\documentclass[final,3p,times,twocolumn]{elsarticle}

\usepackage{graphicx}
\graphicspath{{figures/}{figures/1column/}{figures/2column/}}
\usepackage{subfig}

\usepackage[table]{xcolor} 
\usepackage{color} 
\usepackage[amssymb]{SIunits}

\usepackage{amssymb}
\usepackage{amsthm}
\usepackage{amsmath}

\biboptions{sort&compress}

\journal{Physica A}

\begin{document}

\begin{frontmatter}



\title{Exploring the Mobility of Mobile Phone Users}


\author[a1,a2]{Bal\' azs Cs.\ Cs\'aji}
\author[a3]{Arnaud\ Browet}
\author[a3]{V.A.\ Traag\corref{cor1}}
\ead{vincent.traag@uclouvain.be}
\author[a3,a4,a6]{Jean-Charles\ Delvenne}
\author[a3]{Etienne\ Huens}
\author[a3]{Paul\ Van Dooren} 
\author[a5]{Zbigniew\ Smoreda}
\author[a3]{Vincent D.\ Blondel}

\cortext[cor1]{Corresponding Author}
\address[a1]{Department of Electrical and Electronic Engineering, University of
Melbourne, Australia}
\address[a2]{Computer and Automation Research Institute (SZTAKI), Hungarian
Academy of Sciences}
\address[a3]{ICTEAM Institute, Universit\'e catholique de Louvain, Belgium }
\address[a4]{Namur center for complex systems (NAXYS), Facult\'es Universitaires
Notre-Dame de la Paix, Belgium}
\address[a5]{Sociology and Economics of Networks and Services Department, Orange
Labs, France}
\address[a6]{Center for Operations Research and Econometrics (CORE), Universit\'e
catholique de Louvain, Belgium}

\begin{abstract}

Mobile phone datasets allow for the analysis of human behavior on an unprecedented scale. The social network, temporal dynamics and mobile behavior of mobile phone users have often been analyzed independently from each other using mobile phone datasets. In this article, we explore the connections between various features of human behavior extracted from a large mobile phone dataset. Our observations are based on the analysis of communication data of 100000 anonymized and randomly chosen individuals in a dataset of communications in Portugal. We show that clustering and principal component analysis allow for a significant dimension reduction with limited loss of information. The most important features are related to geographical location. In particular, we observe that most people spend most of their time at only a few locations. With the help of clustering methods, we then robustly identify home and office locations and compare the results with official census data. Finally, we analyze the geographic spread of users' frequent locations and show that commuting distances can be reasonably well explained by a gravity model.

\end{abstract}


\end{frontmatter}


\section*{Abstract}

Mobile phone datasets allow for the analysis of human behavior on an unprecedented scale. The social network, temporal dynamics and mobile behavior of mobile phone users have often been analyzed independently from each other using mobile phone datasets. In this article, we explore the connections between various features of human behavior extracted from a large mobile phone dataset. We show that clustering and principal component analysis allows for a significant dimension reduction with limited loss of information. The most important features are related to geographical location. In particular, we observe that most people spend most of their time at only a few locations. With the help of clustering methods, we then robustly identify home and office locations and compare the results with official census data. Finally, we analyze the geographic spread of users' frequent locations and show that commuting distances can be reasonably well explained by a gravity model.

\section{Introduction}
Information and communication technologies have always been important sources of data and inspiration in sociology, especially in recent decades. These technologies influence the behavior of people, which is a subject of study in itself (e.g.\cite{BALL1968,DEBAILLENCOURT-ET-AL2007,LICOPPESMOREDA2005}), but  they also provide massive amounts of data that can be used to analyze various aspects of human behavior. 

Telephone and mobile phone data have already been used to study social networks,
sometimes in conjunction with features such as gender and age
\cite{SMOREDALICOPPE2000}. More recently, the mobile phone data available to
researchers have been enriched with geographical information. This allows to
analyze regularities, or even
laws\cite{Brockmann2006,GONZALEZETAL-NATURE-2008,Candia2008}, governing the
highly predictable mobility~\cite{Song2010} in everyday life. These insights can
be vital in emergency situations~\cite{Bagrow2011a}, or in (preventing)
spreading of diseases~\cite{Bajardi2011,Balcan2009,Rocha2011} or mobile
viruses~\cite{Wang2009a}. Furthermore, users' mobility and their social network
are intertwined: the one could be used to predict the other~\cite{Wang2011,
Crandall2010}, and the probability of two people calling over a distance follows
a gravity like model~\cite{Lambiotte2008a,Krings2009,Calabrese2011,Levy2010}.
Research has also shown there are geographical clusters of highly connected
antennas~\cite{Expert2011} (e.g. resembling provinces) as well as clusters in
the social network consisting of groups of well connected
people\cite{Palla2007,Blondel2008,Onnela2007}, although the connections between
the two are not yet fully understood~\cite{Onnela2011}. Similar results have
also been obtained in a virtual mobility setting~\cite{Szell2011}.

In this paper, we analyze anonymized communication data from a telecom operator in Portugal. The data cover a period of 15 months and the following information is available for each communication: the times of initiation and termination, the users involved, and the transmitting and receiving antennas (at the beginning of the communication). In addition, we also know the locations (longitude and latitude) of all antennas.

We first present a statistical analysis of the data. We define a set of 50 general features that we compute for each user, and using principal component analysis and clustering methods, we show that these features are highly redundant: they can all be recovered, with a loss of accuracy of less than  $5\%$, using  a reduced set of only five meta-features. 

Observing that the most important features are geographical, we then pay specific attention to the most common locations of each user. By developing a procedure to extract these frequent positions, we observe that people spend most of their time in only a few locations. We then cluster the different calling patterns for each user and each location, and from this, we observe that only two types of locations are clearly identifiable, namely home and work. We compare our results to census data obtained from the Portuguese National Institute of Statistics.

Finally, we analyze in more detail the behavior of users who have exactly one
home location and one office location. This allows us to predict the number of
commuters between different regions of the country using a gravity model. More
precisely, we observe that two different regimes exist, the first involving
distances smaller than \unit{150}{\kilo\meter} (which is half the distance
between the two largest cities) and the second involving larger distances. In
the latter case, only the number of offices in the destination region is
statistically significant.

The fundamental contribution of this paper is that we improve the understanding
of frequent locations. Building on previous location inference
work\cite{ZANG2010}, we construct a method for rigorously determining the type
of locations. It had already been observed that people have only a few top
locations \cite{GONZALEZETAL-NATURE-2008}, but it remained unclear what type of
locations they represent. Although it is often (tacitly) assumed they represent
home and office (\cite{GONZALEZETAL-NATURE-2008, Song2010}), this had never been
rigorously analyzed. We confirm this hypothesis, and also conclude that these
are the only type of locations that are robustly detectable in the data.

\section{Data Mining and Feature Analysis}  \label{sectanalysis}
In this section, we analyze the calling and geographic behaviors of mobile phone users based on features that summarize these behaviors. These features allow us to investigate interdependencies between characteristics such as call durations, the distances of calls, the distances of movements and the frequency of calls. This can be achieved, for example, by analyzing {correlations} between these features. In this section we also use principal component analysis and cluster analysis to better understand these interdependencies.

\subsection{Preprocessing} \label{preprocsec}
Before proceeding with the analysis, some preprocessing of the raw data was necessary. The most important preprocessing step was the application of a {moving weighted average} filter on the calling positions of the users.

This filtering was crucial because the position of the antenna does not always accurately reflect the actual position of the user. Moreover, due to noise (such as that introduced by reflection and scattering in urban environments), the closest antenna is not always the one serving the call. Without proper filtering, these inaccuracies tend to accumulate, particularly for measures such as the total distance traveled.

The filtering was computed as follows. The positions were smoothed independently for all users. Assume that a user made calls at times $t(1), \dots, t(n)$ and the coordinates of the antennas that served the calls are $x(1), \dots, x(n)$. The smoothed positions of the user, denoted by $y(1), \dots, y(n)$, can be calculated as
\begin{equation}
y(i)= \!\! \sum\limits_{j \in B_{\delta} (i) } \!\!\! w(j)\, x(j)
\end{equation}
with $B_{\delta}(i) = \left\{ \, j : \left| t(j) - t(i) \right| \leq \delta \,
\right\}$ where $B_{\delta}(i)$ denotes the indices of those calls that were initiated or received within a maximum interval $\delta$ from the current time of the filtering. We used $\delta=30$min for the dataset. Positions that were further distant from the current time of the decision had proportionally smaller weights:

\begin{equation}
w(j) = 1 - \frac{\left|t(j) - t(i)\right|}{\delta},
\end{equation}
where $i$ denotes the current index of the call that should be smoothed.

In addition to filtering, we took into account those customers who had made and/or received at least $10$ calls during the period analyzed ($15$ months). Moreover, for compression and cluster analysis, we {normalized} (scaled) and {centered} the data. Most of the analysis was performed on $100\,000$ randomly (uniformly) selected users. We performed Student's {t-tests} to examine the statistical significance of the results.

\subsection{Features}
We defined 50 features to summarize users' behavior. Each feature represents one particular aspect of users' behavior as a single number, such as the number of incoming or outgoing calls, the number of people who called or were being called by the user, the position (coordinates) of the user (mean and deviation), the coordinates of the two most frequently used antennas, the durations of incoming or outgoing calls (mean and deviation), the distances of the incoming or outgoing calls (mean and deviation), the directions of the incoming or outgoing calls (mean and deviation) and various movement measures.

\begin{figure}
  \begin{center}
  \includegraphics{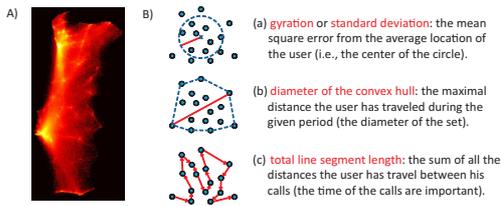}
\end{center}
\caption{(a) The distribution of the average locations of the users. Brighter colors indicate areas in which higher numbers of users have their average locations. (b) Three methods of measuring customer movements: (1) gyration, (2) diameter of the convex hull and (3) total line segment length.}
\label{fig2}
\end{figure}

Individual features themselves can contain considerable information. For example, by analyzing the average locations of the users, we obtained information on the distribution of users across the country, and large cities can be recognized as bright spots in the left panel of Fig.~\ref{fig2}.

In addition to a measure referred to as gyration~\cite{GONZALEZETAL-NATURE-2008}, we propose two additional measures of customer movements: the {diameter of the convex hull} and the {total line segment length}. All three of these measures rely on the positions of the user during calls to give some indication of how much or how far he has traveled. We take into account both incoming (received) and outgoing (initiated) calls. The sequence of the positions (calls) is not important for the first two measures, but it is significant for determining line segment length. This is illustrated in the right panel of Fig.~\ref{fig2}.

Gyration~\cite{GONZALEZETAL-NATURE-2008} measures the deviation (mean square-error) of each of the user's positions from his average location. The diameter of the convex hull measures the maximum distance  between any two positions of the user during a given period. The {total line segment length} sums all of the distances between each pair of consecutive positions of the user. Note that the filtering procedure explained earlier can have a large impact on this final measure.

\subsection{Correlation Analysis}

After computing the values of each feature for each user, we analyzed the {interdependencies} between these features using a {correlation analysis}. As mentioned earlier, we considered $100\,000$ randomly selected users. We used t-statistics to confirm that our results are also valid for the complete dataset. In some cases, the correlations are better analyzed on a {logarithmic} scale, and we have therefore also analyzed the logarithmic correlations.

\begin{table}
\caption{Selected Results of Correlation Analysis}
\begin{center}
\begin{tabular}{|l   l| c c |} \hline
{\em Feature A} & {\em Feature B} & {\em Cor.} & {\em LogCor.}\\
\hline
\hline
No Calls         & No Callers       & .91 & .90\\
Diam Conv Hull & No Antennas     & .55 & .20\\
Avg Duration     & Avg Distance     & .31 & .64\\
No Antennas    & No Calls         & .60 & .68\\
Diam Conv Hull & Avg Duration    & .05 & .18\\
Line Segm Len & No Antennas     & .45 & .75\\
Gyration        & Std Dev Dist   & .60 & .40\\
\hline
\end{tabular}
\end{center}
\label{Corr-table}
\end{table}

Table \ref{Corr-table} shows some of the correlations for the $100\,000$
randomly selected users. The data in this table shows that movement-related features are correlated with some but not all of the other features. Some pairs of features there, such as the number of calls and the number of callers, are highly correlated as expected. Other pairs of features, such as the diameter of the convex hull and the average duration of a call, exhibit weaker correlations. Note that the correlation measures only the {linear} dependencies between two features, and more detailed relationships might be uncovered using more complex methods. We do not pursue this further here, but instead turn to an analysis of the redundancy of the data.

\subsection{Principal Component Analysis}

\begin{figure}
\begin{center}
  \includegraphics{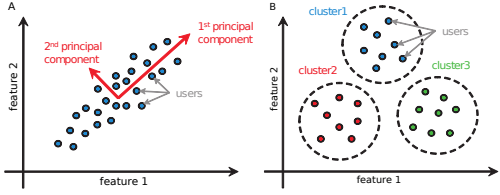}
\end{center}
\caption{Illustration of the basic concept behind (a) principal component analysis and (b) cluster analysis.}
\label{fig4}
\end{figure}

We now analyze the interdependencies between features using another approach. To
what extent are the analyzed features redundant? In other words, how much of the
information represented by one feature can be expressed by (a combination of)
other features? To address this question, we used {principal component analysis}
(PCA), which is widely used in various disciplines.  The basic goal of PCA is to
reduce the dimensions of the data. It can be proven that PCA provides an optimal
linear transformation for mean-square-based dimensionality reduction
\cite{Jolliffe}.

The core idea of PCA is as follow. Let $x_1, \dots, x_n \in \mathbb{R}^{d}$ be
(independent) realizations of a random vector $X$, where we assume that
$\mathbb{E}[X]=0$ (which can be guaranteed, for example, by substracting the
sample mean from the measurements). In our case, the vectors $(x_i)_{i=1}^n$
represent users, while each entry corresponds to a feature. 

We aim at finding {\em orthonormal} vectors $w_1, \dots, w_d \in
\mathbb{R}^{d}$, called the principal components, with the property that
for all $k \in \{1,\dots, d\}$, the linearly transformed vector $Y_k =
W_k^\mathrm{T} X$, where $W_k$ is $[w_1, \dots, w_k]$, explains the maximum
possible variance of $X$. In other words, if we transform our dataset $D = [x_1,
\dots, x_n]$ by $S_k = W_k^\mathrm{T} D$, then we can reconstruct the matrix $D \in
\mathbb{R}^{d \times n}$ from the matrix $S_k \in \mathbb{R}^{k \times n}$ (using
$W_k$) with the smallest possible mean square error. Note that sometimes the
rows of $S_k$, i.e., $s_i = w_i^\mathrm{T} D$ are called the principal
components and the $w_i$'s are referred to as loadings or coefficients.

\newcommand{\argmax}{\mathop{\vphantom{\max}\mathchoice
  {\hbox{arg\,max}}
  {\hbox{arg\,max}}{\mathrm{A}}{\mathrm{A}}}\displaylimits}

A recursive formulation of PCA can be given as follows. Let
\begin{align}
w_1 &= \argmax_{\|w\|=1} \frac{1}{n}\sum_{i=1}^n (w^\mathrm{T} x_i)^2 \nonumber \\
    & \approx \argmax_{\|w\|=1} \mathbb{E} \left[(w^\mathrm{T} X)^2\right] =
      \argmax_{\|w\|=1} \mbox{Var}\left[w^\mathrm{T} X \right].
\end{align}
The vector $w_1$ points toward the direction in which the sample variance of the data is maximized. This is of course an approximation of what we would get using the full (unknown) distribution of $X$. Having defined the first \(k-1\) vectors, the \(k\)-th is determined as
\begin{align}
  w_k &= \argmax_{\|w\|=1} \frac{1}{n}\sum_{i=1}^n
  \Bigg(w^\mathrm{T}\Big(x_i-\sum_{j=1}^{k-1}{w_j w_j^\mathrm{T}
  x_i}\Big)\Bigg)^{\!\!2} \nonumber \\
    &\approx
\argmax_{\|w\|=1,\, w \perp (w_i)_{i=1}^{k-1}}\! \mbox{Var} \left[ w^\mathrm{T}X \right],
\end{align}
which is thus chosen to achieve the highest variance possible while being
orthogonal ($\perp$) to the previous choices.  The vectors $(w_i)_{i=1}^d$ can
be efficiently computed from the (estimate of the) covariance matrix \(\Sigma=
\mathbb{E}\left[X X^\mathrm{T}\right]\), since vector $w_i$, $i \in \{1,\dots,
d\}$, is an eigenvector of the sample covariance matrix corresponding to its
$i$-th largest eigenvalue. The basic concept behind PCA is illustrated in
Fig.~\ref{fig4}.

\begin{table*}
\caption{Compression of Features by Principal Component Analysis}
\begin{center}
\begin{tabular}{|c | c || c | c |} \hline
{\em Variance Kept} & {\em Mean Square Err.} & {\em Dimen.\ Required} & {\em Compress.\ Rate}\\
\hline
\hline
99\% & 1\% & 24 & 52\%\\
\hline
98\% & 2\% & 13 & 74\%\\
\hline
95\% & 5\% & 5 & 90\%\\
\hline
\end{tabular}
\end{center}
\label{PCA-table}
\end{table*}

Our PCA analysis revealed high redundancy among the features analyzed. Table \ref{PCA-table} shows the results of this analysis. It can be seen that if we allow a $1\,\%$ (mean square) error in the variance, the number of features can be reduced by more than $50\,\%$ (from $50$ to $24$). If the allowed error is raised to $5\,\%$, we can further reduce the number of features to $5$, which represents a compression rate of $90\,\%$. In other words, we can build five components using a linear combination of the original features, and using only the values of these five components, we can determine the values of any of the $50$ original features with a $5\,\%$ mean-square error. This implies that the features have many interdependencies and are highly redundant.

In order to identify which features are most relevant, we determined their
importance as follows. PCA produces a set of orthogonal vectors,
$(w_i)_{i=1}^d$, which point toward the directions of maximum variance. As noted
earlier they are eigenvectors corresponding to eigenvalues of the sample covariance
matrix. Furthermore, the $i$-th eigenvalue, $\lambda_i$, equals to the (sample)
variance of $s_i = w_i^\mathrm{T} D$. Then, each original feature can be
identified by an element of the canonical basis. For example, feature 1 can be
identified by $e_1 = \left<1, 0, \dots, 0 \right>^\mathrm{T}$. The {importance}
of feature $i$ can then be defined as the max-norm of the projected vector $e_i$
on the basis defined by $(w_i/{\lambda_i})_{i=1}^{d}$. The basis was thus
scaled to produce larger coordinates in directions of higher variances. Note
that we have also scaled the scores such that the most important feature has a
relative importance of $1$.

\begin{figure*}
\begin{center}
  \includegraphics{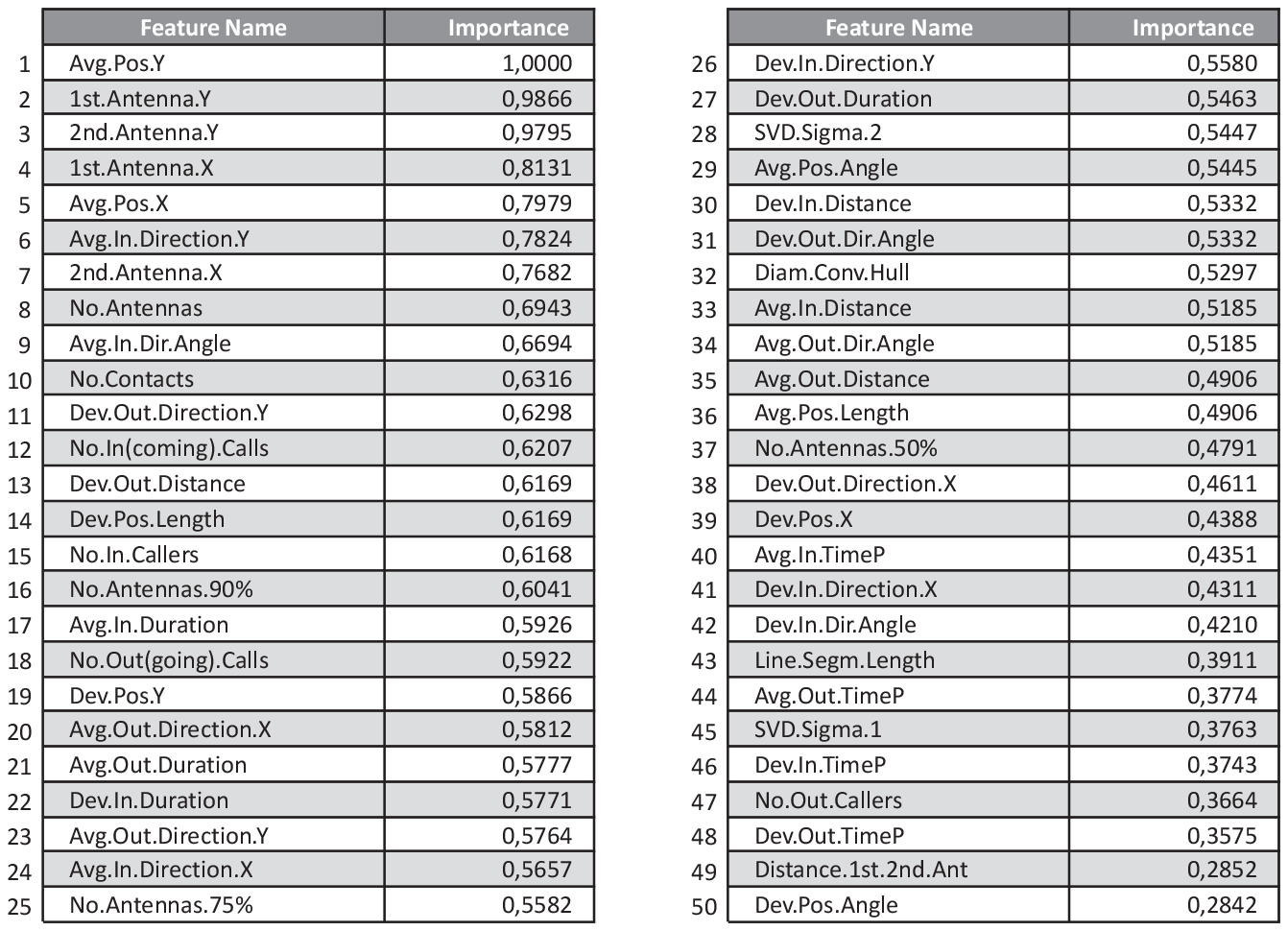}
\end{center}
\caption{The relative importance of each feature according to PCA, defined as the max-norm of the projection of the feature basis on the principal components.}
\label{fig3}
\end{figure*}

Fig.~\ref{fig3} presents a list of the features in order of importance. The most
important features, such as the average position of the user and the coordinates
of the two most often used antennas, are geographic features. This indicates
that the locations of the users and their calls are among the most important
characteristics.

\subsection{Cluster Analysis}
After analyzing the data using PCA, we performed {cluster analysis}, to identify typical user classes based on calling behaviors. We used the {subtractive} clustering method~\cite{Chiu} illustrated in Fig.~\ref{fig4}, which is a variant of the classical mountain method. An advantage of the subtractive clustering method is that it can identify the number of clusters required.

\begin{figure*}
\begin{center}
  \includegraphics{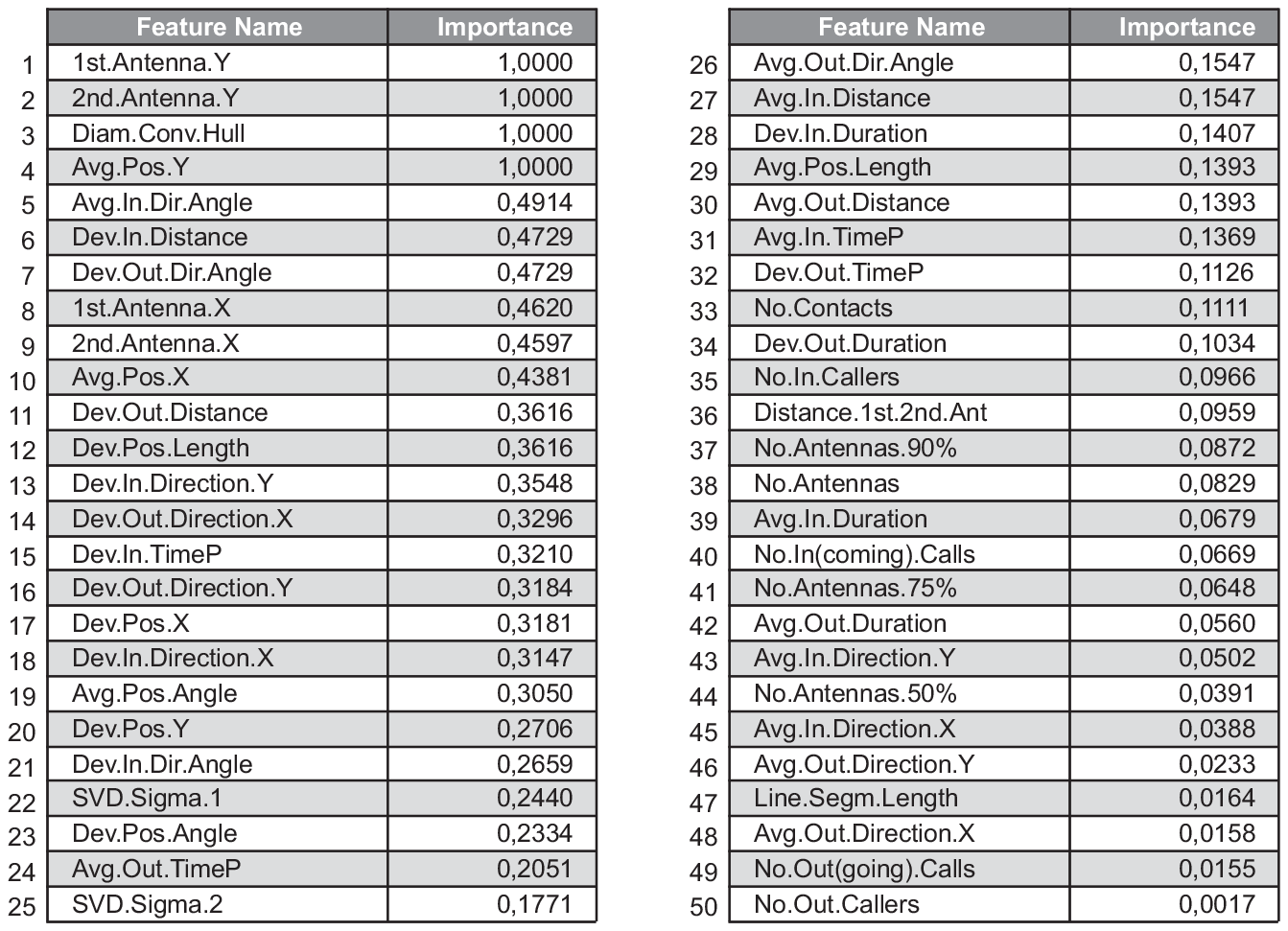}
\end{center}
\caption{The relative importance of each feature according to cluster analysis.}
\label{fig5}
\end{figure*}

The application of subtractive clustering to the {normalized} data for
$100\,000$ uniformly selected customers resulted in $5$ clusters. Each of these
clusters is identified by its {central} element (a vector of feature values) and
its range of influence. As with the PCA, we wished to identify the main
constituent features of these clusters. We therefore performed a similar
analysis as for the PCA, using the vectors of the cluster centers as the basis
for the dominant feature subspace. The results of this ordering, presented in
Fig.~\ref{fig5}, indicate that location- and movement-related features are important characteristics, similar to PCA.

Note that both PCA and cluster analysis reveal clear differences in importance
between the $x$ and $y$ coordinates. This can be expected for an elongated
country such as Portugal and is probably aggravated by the fact that most people
live along the coast.

Although the features concerning the $y$ coordinates have a similar importance
in both the PCA and cluster analysis, there are some clear differences as well.
In general, an explanation of this could be that PCA focuses on global
characteristics, it tries to build components (by linear combination of feature
vectors) which can explain the dataset with minimal mean square error.
Clustering, however, concentrates on local similarities, and tries to find
clusters in which the feature vectors are ``close'' to each other.
Nevertheless, various geographical features have key importance according to
both orderings. The most notable difference concerns the diameter of the convex
hull, which has a very high importance in clustering, while it has a relatively
low importance in PCA. From the PCA analysis this implies that the variance in
the diameter of the convex hull is not important for explaining a large part of
the data. From the cluster analysis, the differences in the diameter of the
convex hull are important, even though the variance might not contribute that
much.  This suggests an interesting effect of the diameter of the convex hull.
Besides the obvious importance of the $y$ position when clustering people, the
diameter of the convex hull separates people that share similar $y$ positions.
In conclusion, the features that are important for clustering people are: (1)
first antenna; (2) second antenna; (3) diameter of convex hull; and (4) average
position.

\section{Frequent Locations} \label{sectarnaud}

In the previous section, we analyzed several features and concluded that the
most important ones are related to geography. Additionally, we observed that
most people spend most of their time in only a few locations. In this section,
we focus on characterizing these frequent locations for each user by analyzing
weekly calling patterns. Once these frequent locations are characterized, we
analyze them in greater depth. A related, although different, concept of
habitats~\cite{Bagrow2012} was recently introduced, where habitats are clusters
of the associated Markov mobility network. However, a single habitat might
contain several frequent locations.

As explained in Section \ref{preprocsec}, the data are noisy, and often, any one of multiple antennas can be used to make a call from a given position. Because this can be true for frequent locations such as home and the office, we first develop a method for estimating which antennas are relevant for characterizing such locations.

After extracting the frequent locations for each user, we estimate these positions more precisely using a maximum likelihood approach. We then present various statistics using these estimated positions. In particular, we estimate the amount of time people spend at work and home, characterize different combinations of frequent locations (multiple `homes' or `offices'), estimate the geographical density of homes and offices, compare our estimates to independent statistics and, finally, analyze distances between home and office (commuting distances).

\subsection{Detection of Frequent Locations}

Detecting the most common locations of a user is only possible if enough calls involving that user are recorded\footnote{In this section, we include both calls and text messages because we want to maximize the information on antenna usage; we refer to both as ``calls''.}. For users who make only a few calls, no locations can be called ``frequent'' with any certainty. We therefore selected only users who make at least one call a day on average and who make consecutive calls within $24$ hours 80\% of the time. The latter constraint requires a certain regularity of users, and excludes users with highly bursty behavior\cite{GONZALEZETAL-NATURE-2008}. From this selection, we selected a random sample of $100\,000$ users.

For detecting frequent locations, it is appropriate to begin by identifying the most frequently used antenna (MFA). However, as stated earlier, the same antenna is not always used for calls made from a given position (due to load balancing or the effects of noise on the signal). Hence, other antennas located near the MFA may also be used to serve the frequent location. We must therefore consider sets of antennas that are relatively close together.

We first performed a Voronoi tessellation, which partitions the space into cells based on the distance between each point and the closest antenna. Each Voronoi cell includes the set of points that are closer to the antenna located in that cell than to any other antenna. Based on the Voronoi tessellation, a graph can be created in which nodes are neighbors if their associated Voronoi cells are adjacent. Each node corresponds to an antenna, and its neighbors are called the Delaunay neighbors.

We next grouped antennas around the MFA based on Delaunay neighborship. More precisely, we defined the Delaunay radius of each antenna to be the largest distance between an antenna and any of its Delaunay neighbors (this is later used in the estimation of the position; see Section~\ref{sec:Neighborhood} for more information). We then merged all antennas around the MFA that are within twice this radius\footnote{We observe that taking twice the Delaunay radius yields an error of less than $0.1$\% for estimating positions. See Section~\ref{sec:Neighborhood} for more information.} and assigned them one ``location'', the position of which will be defined later.

After identifying the first MFA and merging the surrounding antennas, we moved on to the remaining antennas, selecting the most frequently used of those and repeating the procedure described above. We continued iterating until we identified a set of antennas that represented less than 5\% of a  user's calls. We repeated this for each user in our selection and thus obtained a number of frequent locations for all users.

\begin{figure}
  \begin{center}
     \includegraphics{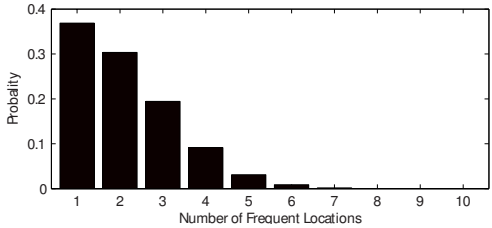}
  \end{center}
  \caption{Histogram of the number of frequent locations per user}
  \label{fig:histFL}
\end{figure}

The results of this procedure are summarized in Fig.~\ref{fig:histFL} and indicate that the average number of frequent locations per user is approximately $2.14$ and that $95\%$ of the users have fewer than $4$ frequent locations. This implies that the 3 or 4 most common locations are sufficient to predict the position of user, most of the time~\cite{Song2010}. A substantial number of users have only one single frequent location, which is usually an office or a home location (as we will see later on). This could reflect the possession of separate business and private phones, one of which is (almost) exclusively used at work and the other only at home.

\subsection{Clustering of Weekly Calling Patterns}

\begin{figure}
  \begin{center}
   \includegraphics{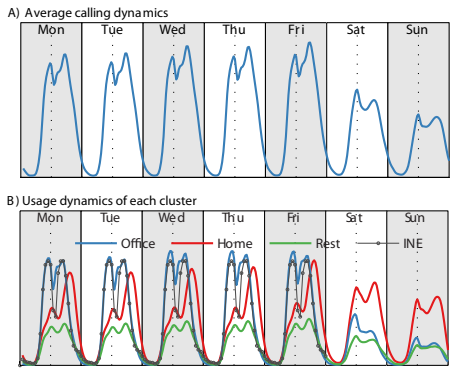}
  \end{center}
  \caption{Weekly dynamics on average (a) and for the three clusters (b) detected using k-means clustering: home, office and the remainder, shown along with independent time usage statistics from the Portuguese National Institute of Statistics (INE)~\cite{INE1999}. Using more clusters yields similar results. The dotted lines indicate noon of each day.}
  \label{fig:home_office}
  \label{fig:avg_pattern}
\end{figure}

The data show two clearly identifiable periodic dynamics in mobile phone use: a daily cycle and a weekly cycle, as illustrated in Fig.~\ref{fig:avg_pattern}. The daily cycle largely follows the human circadian rhythm, with a clear drop in activity during the night, a gradual increase in the morning and a decrease in the evening, with a small dip around lunch time. The weekly dynamic is related to the workweek, with different behavior on weekends as compared to work days.

We collected all of the calls made using antennas associated with each frequent location. Because we have the time stamps (beginning and end) of each call, we know the times at which each frequent location is used. The description of this usage at the weekly scale seems to be especially suitable for further analysis. We therefore divided the week into 168 hours and aggregated the usage pattern of the whole period. This resulted in a $168-$dimensional vector per frequent location with the calling frequency for one hour in each entry.

Based on the aggregated call vectors for all frequent locations, we performed k-means clustering. We ran this clustering for $k=\{2,\ldots,10\}$ to investigate what patterns of usage could be distinguished. We found that using $k=3$ yielded clear results, as displayed in Fig.~\ref{fig:home_office}. The first cluster clearly represents a pattern related to work. During the weekdays, an increase in the usage of these antennas occurs during the morning, followed by a small dip around noon, and a decrease in usage from around 6 p.m. on. During the weekend, these antennas are used far less. This pattern is in excellent agreement with independent statistics from the Portuguese National Institute of Statistics (INE) in terms of time spent at work, as shown in Fig.~\ref{fig:home_office}. The second cluster reflects a pattern of usage that appears to be more closely associated with a home position. The usage of these antennas is lower during the day, and the maximum usage occurs during the evening. These antennas are also used more during the weekend than are the antennas in the first cluster. Finally, the third cluster appears simply to contain locations that do not follow the dynamics of the previous two clusters. This cluster follows the more general dynamic displayed in Fig.~\ref{fig:avg_pattern}.

We observed that when more than three clusters are considered, they tend to yield results very similar to those shown here. We expected that we would be able to identify additional patterns of usage, such as those of calls made by students with a different rhythm from working people or calls made from weekend houses that show no activity during the week, but we did not observe these patterns. Such patterns certainly do exist, but they appear to be marginal when compared to the established home and office routine. Hence, there appears to be no identifiable patterns of usage other than the home and office patterns described above. However, using only two clusters obfuscates this result, and the separation between home and office positions is less clear in this case.

\begin{table}
\begin{center}
\rowcolors{2}{gray!25}{white}
\begin{tabular}{cccc}
    \rowcolor{white}
$\#$ Home & $\#$ Office & $\#$ Unidentified & \%\\
\hline \hline
 &  1&   &  16.8\\ 
1&   &   &  16.5\\
1&   &  1&  9.1\\
1&  1&   &  6.6\\
 &  1&  1&  6.0\\
1&   &  2&  5.0\\
 &   &  1&  3.5\\
1&  1&  1&  3.5\\
 &   &  2&  3.4\\
 &  1&  2&  2.7
\end{tabular}
\caption{The $10$ most frequent combinations of frequent locations. Each
  combination is composed of the number of homes, offices and unidentified
  locations a user has. Each row indicates such a combination. The empty entries
  indicate no such type of location is present in a combination. The last column
  contains the percentage of how often such a combination occurs.}
\label{locPattern}
\end{center}
\end{table}

The top 10 most frequent combinations of frequent locations are displayed in Table~\ref{locPattern}. Approximately 32\% of the users have either a single home location or a single office location alone, whereas only 3.5\% have only a single unidentified location. For users with two frequent locations, the most common combination is one home location and one unidentified location. Only 6.6\% of all users have the combination of one home location, one office location and no unidentified locations. Approximately 85\% of the users have at most one home and/or one office location, and approximately 12\% of the users have exactly one home and one office location (and possibly multiple unidentified locations).

\begin{figure}
  \begin{center}
     \includegraphics{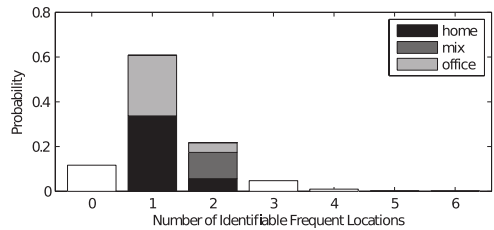}
  \end{center}
  \caption{Histogram of the number of frequent locations per user, with visual separation between ``home'' locations, ``office'' location and mixed locations (i.e., users with one of each)}
  \label{fig:histFLID}
\end{figure}

Of all frequent locations, approximately $60\%$ ca be classified as ``home'' or
``office'' (as in the first two columns of Table~\ref{locPattern}). We observed
that users tend to have no more than two identifiable positions, as depicted in
Fig.~\ref{fig:histFLID}. The majority of users have only one identifiable location, which is by definition either home or office. For users with two identifiable locations, over 50\% have both a home and an office, and the rest has either two homes or two offices.

\subsection{Estimating the Position of Frequent Locations}

  \subsubsection{Basic model}

  We propose a model to estimate the position of the home, the office and other frequent locations. We consider a simplified version of the model proposed in \cite{ZANG2010}, which was also used in~\cite{SocialEvent2011}. The underlying idea is that users connect to the antenna that has the highest signal strength, which is not necessarily the closest antenna. 

\begin{figure}
  \begin{center}
     \includegraphics{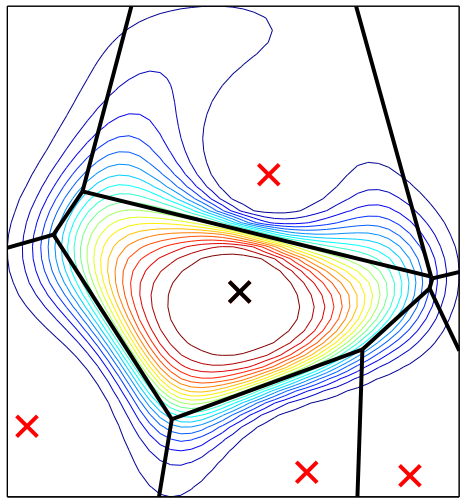}
  \end{center}
  \caption{Probability density $\Pr(a=i | x)$ (represented by topographic curves) for a particular antenna $i$ (central black `X'), with neighboring antennas (red `X's) and the local Voronoi tessellation (dark lines) also shown. The probability density can be seen as a smoothed Voronoi tessellation in which there is a small probability of connection to antenna $i$ when the user is in another Voronoi cell.}
  \label{fig:antenna_dens}
\end{figure}

  We begin by estimating the total signal strength of an antenna $i$ at a
  certain position $x$. We assume, similarly to~\cite{ZANG2010}, that the total
  signal strength consists of three components: the power of the antennas,
  the loss of signal strength over distance and some stochastic fading of the signal due to scattering and reflection in the environment. Specifically, we use the following parameters.
  \begin{itemize}
    \item The position of antenna $i$ is denoted by $X_i$.
    \item The power is denoted by $p_i$, and we assume this to be constant and equal for all antennas because we have no information regarding the power of the antennas. Therefore, $p_i=p$ for all $i$.
    \item The loss of signal at position $x$ for antenna $i$ is modeled as
      \begin{equation}
        L_i(x) = \frac{1}{\| x - X_i \|^\beta},
      \end{equation}
      where $\beta$ is a parameter indicating how quickly the signal decays.
    \item The so-called Rayleigh fading of the signal from antenna $i$ can be modeled by a unit mean exponential random variable $R_i$~\cite{Tse:2005p10323}, for which the cumulative distribution function (cdf) is
      \begin{equation}
        \Pr(R_i \leq r) = F(r) = 1 - e^{-r}.
      \end{equation}
      Furthermore, we assume all $R_i$ to be independent.
  \end{itemize}
  The total signal strength $S_i(x)$ of antenna $i$ at location $x$ is then modeled as 
  \begin{equation}
		S_i(x) = p L_i(x) R_i,
	\label{eq:signalstrength}
	\end{equation}
	and we model the probability that a user at position $x$ connects to antenna $i$, $\Pr(a=i | x)$, as the probability that the signal strength of antenna $i$ is larger than that of any other antenna:
	\begin{align}
		\Pr(a=i | x) &= \Pr(S_i(x) > S_j(x), ~~ \forall j) \nonumber \\
                 &= \prod_j{\Pr(S_i(x) > S_j(x))}.
		\label{equ:probabilityAiX}
	\end{align}
  This probability density is displayed in Fig.~\ref{fig:antenna_dens}.
  
	This probability can be seen as a smoothed Voronoi tessellation, in which a user will always connect to the closest antenna, by taking the limit of $\beta \to \infty$. In that case, we are essentially considering the situation in which the path loss is dominant over the Rayleigh fading. Hence, little noise is involved, and whenever a closer antenna exists, it will be used.

  \subsubsection{Antenna neighborhoods}\label{sec:Neighborhood}
  
  As mentioned in the previous section, the probability that a user will connect to a specific antenna depends on the position of other nearby antennas.  The relevant set of antennas $\mathcal{X}$ can be rather large, which can slow down the computation of the probabilities. Using a local approximation might accelerate this process without affecting the results.
  
  The idea of using a local approximation is tied to the decreased probability that a call will be linked to an antenna that is far away. Only some of the antennas around a given position are in fact relevant. It therefore seems natural to construct local neighborhoods of antennas so as to make the method more efficient without introducing any significant error.
  
  We define the neighborhood $\mathcal{X}_i$ and the domain $\mathcal{D}_i$ of antenna $i$ to consist of the smallest circle enclosing at least all of the Delaunay neighbors (and possibly more). As mentioned previously, the Delaunay neighbors are those antennas located in adjacent Voronoi cells.
  \begin{itemize}
    \item For each antenna, we select all Delaunay neighbors and then select the
      maximum distance between the focal antenna and any of these neighbors:
      \begin{equation}
        \rho_i = \max \{ d(X_i, X_j) | j \text{~Delaunay neigh. of~} i\},
      \end{equation}
      where $d(X_i,X_j)$ is the distance between antenna $i$ and $j$.
    \item We then define the domain 
      \begin{equation}
        \mathcal{D}_i = \{x | \|x - X_i\| \leq \delta \rho_i\}
        \label{eq:domain}
      \end{equation}
      as the region within radius $\delta \rho_i$, where $\delta $ is a scaling factor. We observe that choosing $\delta =2$ leads to an error of less than $0.1\%$ in the computation of $\Pr(a=i | x)$ compared\footnote{average error based on $1000$ random points} to using the entire set $\mathcal{X}$.
    \item Finally, the set of Delaunay neighbors\footnote{To deal with antennas near the border of the country (for which the Delaunay neighbors can be far away), we take this border into account, and create a slightly different neighbor set.} is taken as all antennas within this region:
      \begin{equation} \mathcal{X}_i = \{j | X_j \in \mathcal{D}_i \text{~for~} j \in \mathcal{X} \}.\end{equation}
    Note that this set contains at least all of the Delaunay neighbors and may also contain other antennas.  
  \end{itemize}
   
  Finally, using equation (\ref{equ:probabilityAiX}), we approximate the probability  as 
  \begin{equation}
    \Pr(a=i | x) \approx \prod_{j \in\mathcal{X}_i}{Pr(S_i(x) > S_j(x))},
    \label{eq:probabilityAisubX}
  \end{equation}
  leading to a large reduction in the computational time required.
	
  \subsubsection{Maximum Likelihood Estimation}
  
  We use the model explained above to more accurately estimate the position of each frequent location. For each such location, we know the number of calls $k_i$ made using antenna $i$. The probability that $k_i$ calls were made using antenna $i$ given position $x$ is then $\Pr(a=i | x)^{k_i}$.  Hence, the log likelihood of observing call frequencies $k$ for the antennas in $\mathcal{X}_f$, where $f$ is the MFA of a frequent location, for a certain position $x$ is
  \begin{equation}
    \log \mathcal{L}(x|k) = \sum_{i \in \mathcal{X}_f} k_i \log \Pr(a=i | x).
    \label{equ:log_likelihood}
  \end{equation}
  The maximum likelihood estimate (MLE) $\hat{x}$ of the position of a frequent location is then given by
  \begin{equation}
    \hat{x} = \arg \max_{x} \log \mathcal{L}(x|k).
    \label{equ:mle}
  \end{equation}

To find the MLE, we employ a derivative-free optimization scheme because the
gradient of the likelihood function is costly to evaluate. In particular, we use
the Nelder-Mead algorithm\cite{NelderMead}, initialized with the weighted
average position of the antennas associated with the frequent location. The
distance between the average position of the antennas and the MLE is
\unit{1.7}{\kilo\meter} on average and reaches a maximum of approximately
\unit{35}{\kilo\meter}. This shows that although using the average position provides a reasonable approximation, it is not always accurate.

\subsection{Results}
We now analyze the results of the position estimation. First, we present our results concerning the geographical distribution of frequent locations around the country and compare these results to independent statistics. We then analyze commuting distances, i.e., the distances of travel between home and office, and develop a model of the number of commuters between each pair of counties\footnote{We used the NUTS-3 data defined by Eurostat, which, in the case of Portugal, consists of groups of municipalities; we refer to these as ``counties'' for simplicity.}.

\subsubsection{Population density estimation}

\begin{figure*}
  \begin{center}
   \includegraphics{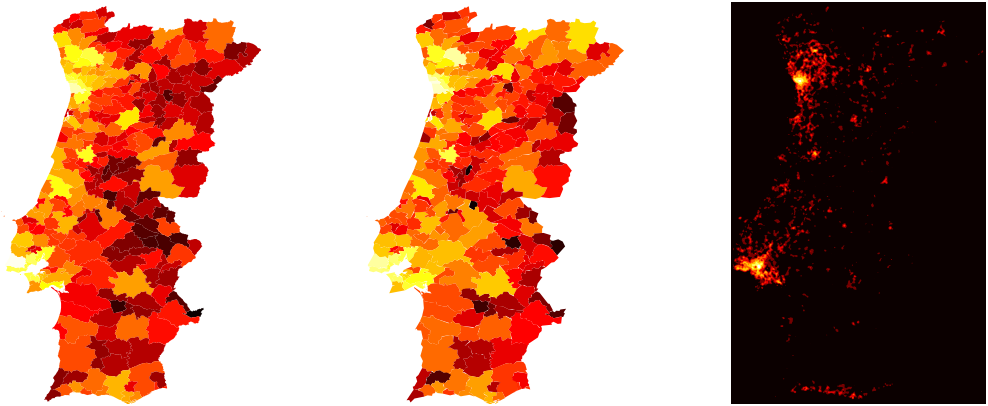}
  \end{center}
  \caption{(a) Population sizes per county throughout the country (based on statistics from INE), (b) estimated number of homes per county, and (c) the distribution of all frequent locations. Lighter colors indicate higher values.}
  \label{fig:distr_FL}
  \label{fig:map_pop_dens}
\end{figure*}

The position estimates of all frequent locations can be used to analyze the population distribution throughout the country. Using the county level data, we counted the number of home locations for each county. We then compared these results to population density data obtained from the \emph{Instituto Nacional de Estatistica\footnote{http://www.ine.pt}} (INE). As shown in Fig.~\ref{fig:map_pop_dens}, there is a strong correspondence between the INE population data for each county and our estimate. The correlation between the two is 0.92. This indicates that we can accurately estimate population size based on the mobile phone data. A more accurate density plot of the frequent locations is shown in Fig.~\ref{fig:distr_FL}, which illustrate that these locations are concentrated in the cities. A comparison of Fig.~\ref{fig:distr_FL} to the distribution of the average positions of users over the entire period (Fig.~\ref{fig2}), shows that the distribution of frequent locations is more pronounced. Average positions are likely to be distorted by commutes and to interpolate between home and office.

\subsubsection{Commuting distances}

\begin{figure}
  \begin{center}
     \includegraphics{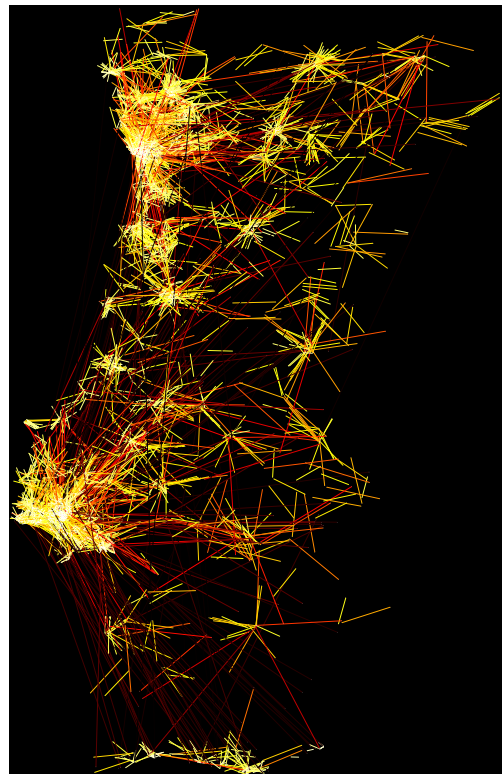}
  \end{center}
  \caption{Commute map for our sample of users. Brighter colors indicate smaller commuting distances. Most of the commutes cover only small distances, although some commutes span half the country. The number of commutes decays approximately log-normally with distance.}
  \label{fig:commute_map}
\end{figure}

The home and office positions determined above can be used to estimate commuting distances. For individuals who have more than one home or one office, multiple commuting distances could be calculated, but it would be unclear which distance is the ``correct'' one. Therefore, for this analysis, we considered only the $12\%$ of users who have exactly one home and one office (and possibly some unidentified frequent locations). This means that each user considered has exactly one commuting distance. These commutes are plotted in Fig.~\ref{fig:commute_map}, with smaller distances indicated in brighter colors. Two things stand out on this map. First, the two largest cities in Portugal, Porto and Lisbon, are clearly discernible. Second, most of the cities appear to predominantly attract people living in the immediate surroundings.

\begin{figure}
    \begin{center}
       \includegraphics{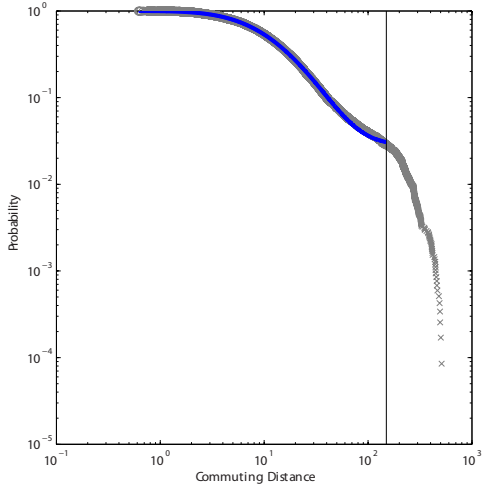}
    \end{center}
    \caption{The distribution of commuting distances revealed by the analysis of
    mobile phone data. These distances exhibit a log-normal distribution for $d
    < \unit{150}{\kilo\meter}$ (blue line). The full distribution is shown, with
    the line at \unit{150}{\kilo\meter} separating the two different regimes.
    These two distinct regimes probably arise because almost all of continental
    Portugal is within \unit{150}{\kilo\meter} of either Porto or Lisbon.}
    \label{fig:commute_dist}
\end{figure}

The distribution of commuting distances depicted in Fig.~\ref{fig:commute_dist}
appears to be affected by the location of Porto and Lisbon. Two different
regimes can be discerned: one regime reflecting commuting distances of less than
\unit{150}{\kilo\meter} and the other reflecting larger distances. This
coincides with the distance between Lisbon and Porto, which is approximately
\unit{300}{\kilo\meter}. In fact, most of Portugal is within
\unit{150}{\kilo\meter} of one of these two cities. This suggests that most
people tend to work no further away than the closest largest city, i.e., it is
unlikely that people living near Porto work in Lisbon. The set of commuting
distances that are less than \unit{150}{\kilo\meter} can be reasonably well fitted using a log-normal distribution with parameters $\mu=2.35$ and $\sigma=0.94$, as displayed in Fig.~\ref{fig:commute_dist}.

A common model for analyzing commuting distance is the gravity model~\cite{Balcan2009, GravityModel}, although recently another parameterless model has been suggested~\cite{Simini2012}. This model formulates the number of trips $w_{ij}$ made between two locations $i$ and $j$ as proportional to the population sizes at the origin $P_i$ and at the destination $P_j$, with some decay, depending on the distance $d_{ij}$ between $i$ and $j$. More precisely, the model is formulated as
\begin{equation}
 \hat{w}_{ij} \sim \frac{P^\alpha_iP^\beta_j}{f(d_{ij})},
\end{equation}
where $f(d_{ij})$ is usually taken as either a power law $d^\gamma_{ij}$ or an exponential decay $e^{\gamma d_{ij}}$, with parameters $\alpha$, $\beta$ and $\gamma$ to be estimated from the data.

Here, we formulate the gravity model in terms of the number of trips (commutes) made between county $i$ and county $j$. Instead of simply considering the population size as $P_i$ and $P_j$, we can take into account our previous calculations of the distributions of both home positions and office positions. The probability of a trip from $i$ to $j$ can then be formulated in terms of the number of home locations at the origin $H_i$ and the number of office locations at the destination $O_j$. 

\begin{table*}
  \begin{center}
    \begin{tabular}{llll}
      Coefficient & Variable & $d_{ij} < \unit{150}{\kilo\meter}$ & $d_{ij} \geq \unit{150}{\kilo\meter}$ \\
    \hline \hline
    $\alpha$ & Number of homes at origin  & $0.17^{**} \pm 0.013$  & $0.018 \pm 0.013$ \\
    $\beta$  & Number of offices at destination & $0.21^{**} \pm 0.013$  & $0.030^{*} \pm 0.012$ \\
    $\gamma$ & Distance  & $0.37^{**} \pm 0.018$  & $0.13 \pm 0.11$ \\
      & $R^2$ & $0.52$ & $0.26$ \\
      & $R^2$ (exponential fit) & $0.46$ & $0.26$
    \end{tabular}
  \end{center}

  \caption{Fitted parameters and $R^2$ of the gravity model with $f(d_{ij}) = d_ij^\gamma$, with standard errors reported. We also report $R^2$ for the exponential fit $f(d_{ij}) = e^{\gamma d_{ij}}$, which is slightly worse. ${}^{**}$ $p < 0.001$, ${}^*$ $p < 0.05$.}
  \label{tab:gravity_fit}
\end{table*}

\begin{figure*}
  \begin{center}
    \subfloat[Power law decay]{\includegraphics[width=0.5\textwidth]{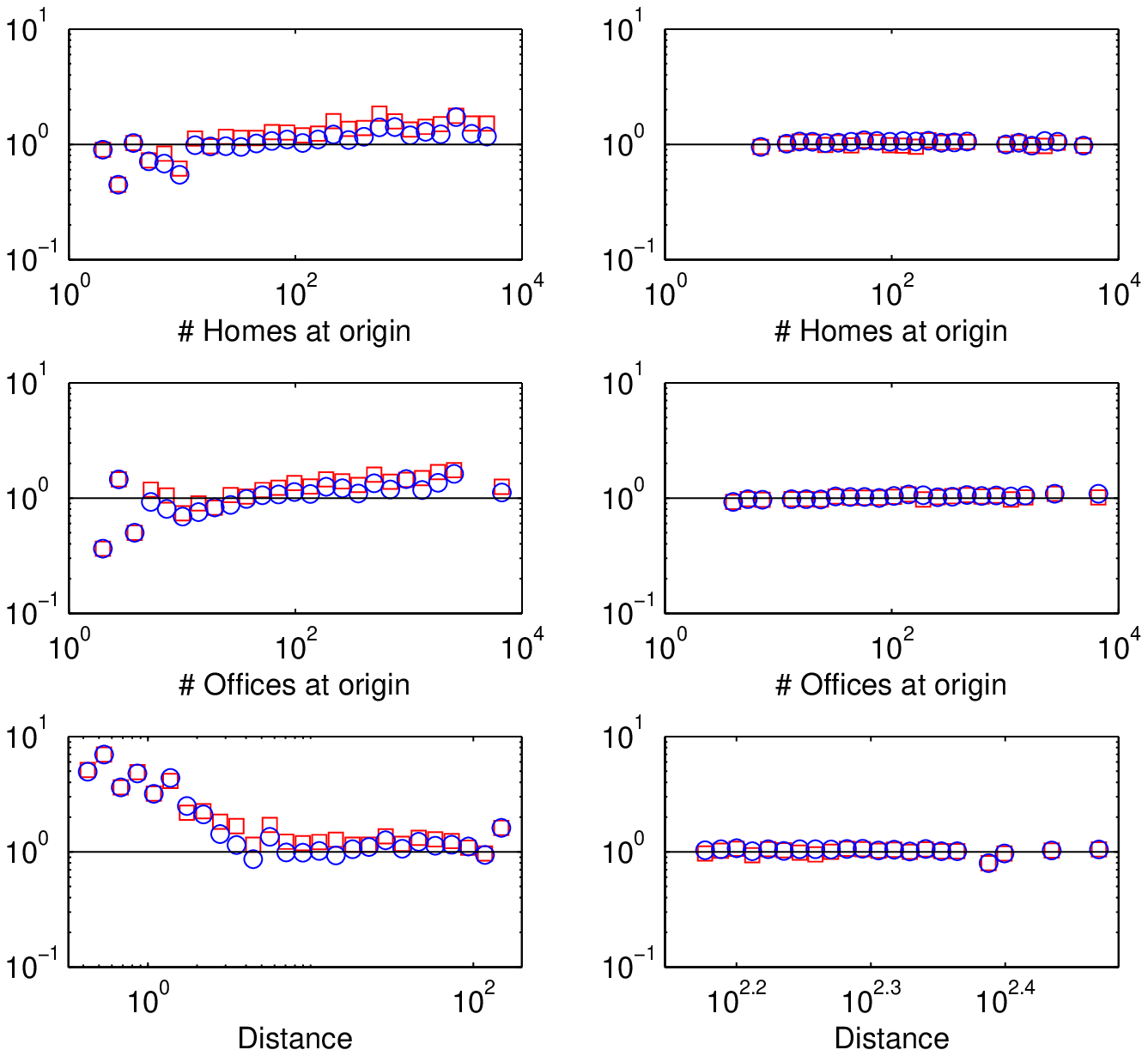}}
    \subfloat[Exponential decay]{\includegraphics[width=0.5\textwidth]{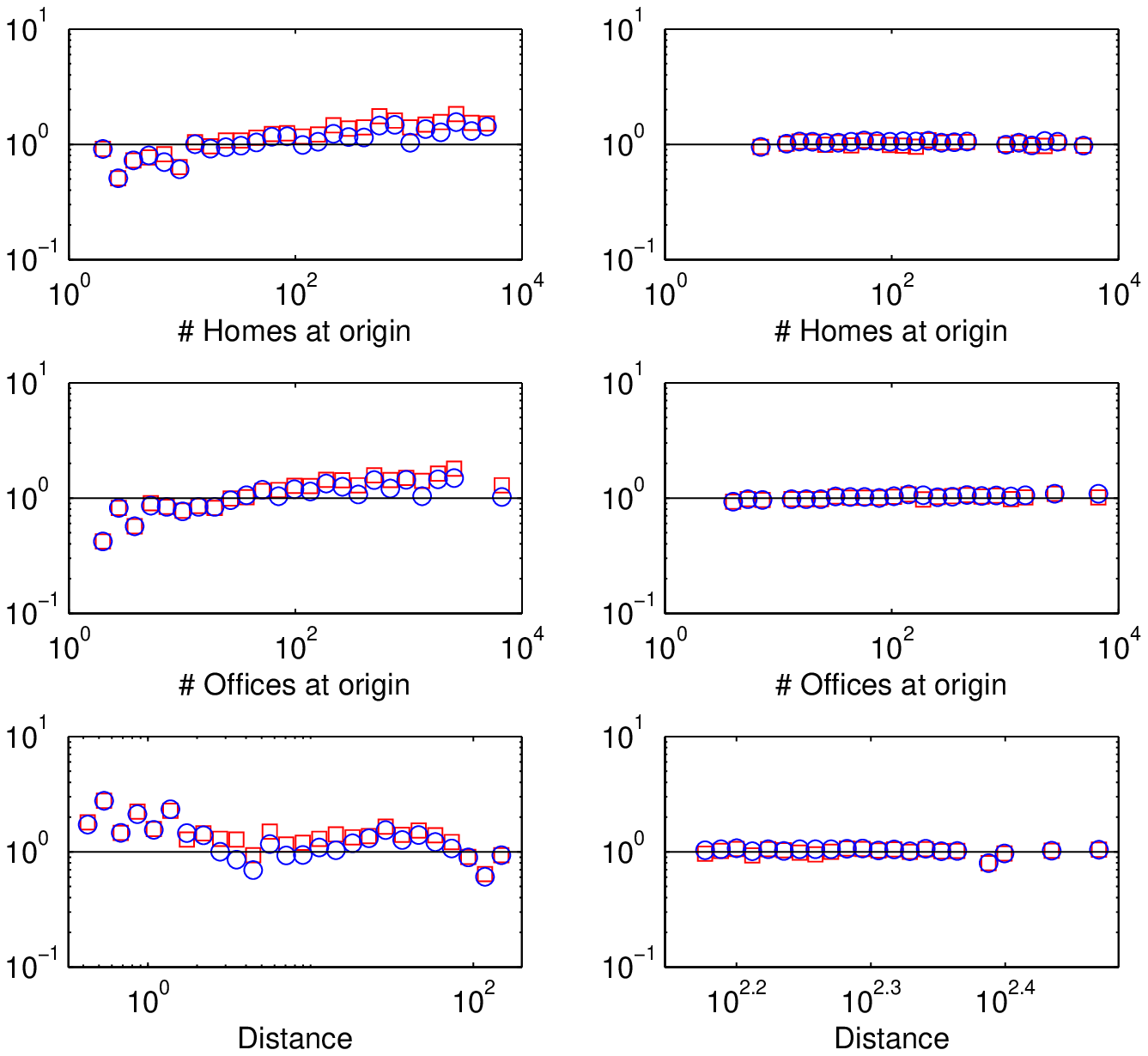}}
  \end{center}
  \caption{Plots of the prediction ratio $\hat{w}_{ij}/w_{ij}$ for commuting
  distance of $d < \unit{150}{\kilo\meter}$ (left panels) and $d_{ij} \geq
  \unit{150}{\kilo\meter}$ (right panels) for (a) the power law decay $f(d_{ij})
  = d_{ij}^\gamma$, and (b) the exponential decay $f(d_{ij}) = e^{\gamma
    d_{ij}}$.  Red squares indicate mean values, and blue circles indicate
  medians.}
  \label{fig:gravity_fit}
\end{figure*}

Again, we discern two regimes: a close-range regime with $d_{ij} < \unit{150}{\kilo\meter}$ and a long-distance regime with $d_{ij} \geq \unit{150}{\kilo\meter}$. Fitting both the power law decay and the exponential decay, we find that the power law decay provides a slightly better fit. The results are displayed in Fig.~\ref{fig:gravity_fit} and in Table~\ref{tab:gravity_fit}. Interestingly, the decay distance parameter $\gamma$ for large distances is not significant, suggesting that for distances $d_{ij} \geq \unit{150}{\kilo\meter}$, the number of trips no longer depends on the actual distance. In fact, the only coefficient that is significant for large distances is the coefficient of the number of offices at the destination. Thus, for larger distances, only the number of work opportunities at the destination appears to be important.

The fit of the model is better when the numbers of home and office locations per county are used than when the population sizes are used. As shown in Table~\ref{tab:gravity_fit}, the values of $R^2$ for the two regimes are $0.52$ and $0.26$, respectively, when the numbers of home and office locations are used, compared to $0.43$ and $0.24$, respectively, when population sizes are used. Hence, it is worth taking into account the numbers of offices and homes when modeling commuting distances instead of simply using population size as an approximation for both. 
In the present case, the model slightly overestimates the number of shorter
commutes, indicating that there is room for improvement. This deviation might be
due to the aggregation of information at a small resolution. On the other hand,
this might also be due to a real effect: distances below some threshold have no
effect. In this case, trips under about \unit{2}{\kilo\meter} should be almost
unaffected by distance. Higher resolution data is needed to investigate this in
more detail.

\section{Conclusion} \label{sectpaul}

In this study, we analyzed the behavior of mobile phone customers based on their calling habits. We first sampled $100\,000$ customers randomly and filtered their locations, as these are based on associated antenna locations, which are subject to disturbances. We then defined and computed $50$ features that describe the calling behaviors of the customers. We performed a correlation analysis on these features, which showed that movement- and location-related features are correlated with many other features. We then analyzed the data using principal component analysis (PCA). This showed that the original features are highly redundant and can be efficiently compressed if some reconstruction error (e.g., $5\,\%$) is allowed. We also performed a cluster analysis and that revealed a small number of typical user classes. We computed the relative importance of each feature in the PCA and the cluster analysis and found that location- and movement-related features are especially important in both cases. We therefore analyzed the users' most common locations.

We clustered these frequent locations based on weekly calling patterns and found that only home and office locations could be clearly identified. Other patterns of usage (such as use from weekend houses) are surely present in the data, but these are marginal when compared to the clear pattern of use from home and office locations. We characterized the number of frequent locations for each user and the most common combinations of frequent locations (e.g., multiple houses or offices). Finally, we estimated the positions of frequent locations based on a probabilistic inference framework. Using these positions, we derived a fairly accurate estimate of the distribution of the population, which showed a correlation of 0.92 with independent population statistics. These positions also allowed us to analyze commuting distances, and we found that the data are reasonably well explained by a gravity model. This model works better when the numbers of homes and offices are considered instead of population sizes. This indicates that when analyzing commuting distances, it is worth taking the distribution of home and office location into account.

The present study represents an exploratory analysis of the data. Further research into the frequent locations and associated user behavior should be undertaken. This data set contains both geographical data and social network data, and it would be interesting to further analyze the interaction between the two.

\section*{Acknowledgments}
The authors acknowledge support from the grant ``Actions de recherche concert\'ees --- Large Graphs and Networks'' of the Communaut\'e Fran\c caise de Belgique and from the Belgian Network DYSCO (Dynamical Systems, Control, and Optimization), funded by the Interuniversity Attraction Poles Programme initiated by the Belgian State Science Policy Office. The funders had no role in study design, data collection and analysis, decision to publish, or preparation of the manuscript.





\bibliographystyle{model1-num-names}
\bibliography{bibliography}







\end{document}